\documentclass[twocolumn,aps,prd,amsmath,amssymb,nofootinbib,superscriptaddress]{revtex4-1}

\pdfoutput=1

\usepackage{graphicx,amsmath,amssymb,feynmf}

\newcommand{\bea}{\begin{eqnarray}}
\newcommand{\eea}{\end{eqnarray}}
\newcommand{\beq}{\begin{equation}}
\newcommand{\eeq}{\end{equation}}

\newcommand{\pw}{\beta}

\newcommand{\f}{\Phi}
\newcommand{\fr}{\phi}
\newcommand{\rhoc}{\rho_c}
\newcommand{\rc}{R_c}
\newcommand{\mc}{M_c}
\newcommand{\nc}{n_c}
\newcommand{\ex}{\alpha}
\newcommand{\F}{F}
\newcommand{\Fa}{F_a}
\newcommand{\N}{N}
\newcommand{\mphi}{\psi}
\newcommand{\mS}{S}

\newcommand{\g}{\kappa}
\newcommand{\coef}{\xi}
\newcommand{\gam}{\gamma}
\newcommand{\ind}{p}
\newcommand{\expo}{a}

\begin{document}

\title{Can Light Dark Matter Solve the Core-Cusp Problem?}

\author{Heling Deng}
\email{heling.deng@tufts.edu}
\affiliation{Institute of Cosmology, Department of Physics and Astronomy,
Tufts University, Medford, MA 02155, USA}

\author{Mark P. Hertzberg}
\email{mark.hertzberg@tufts.edu}
\affiliation{Institute of Cosmology, Department of Physics and Astronomy, 
Tufts University, Medford, MA  02155, USA}

\author{Mohammad Hossein Namjoo}
\email{namjoo@mit.edu}
\affiliation{Department of Physics, Massachusetts Institute of Technology, Cambridge, MA 02139, USA}

\author{Ali Masoumi}
\email{ali@cosmos.phy.tufts.edu}
\affiliation{Institute of Cosmology, Department of Physics and Astronomy,
Tufts University, Medford, MA 02155, USA}

\begin{abstract}
Recently there has been much interest in light dark matter, especially ultra-light axions, as they may provide a solution to the core-cusp problem at the center of galaxies. Since very light bosons can have a de Broglie wavelength that is of astrophysical size, they can smooth out the centers of galaxies to produce a core, as opposed to vanilla dark matter models, and so it has been suggested that this solves the core-cusp problem. In this work, we critically examine this claim. While an ultra-light particle will indeed lead to a core, we examine whether the relationship between the density of the core and its radius matches the data over a range of galaxies. We first review data that shows the core density of a galaxy $\rho_c$ varies as a function of the core radius $R_c$ as $\rho_c\propto 1/R_c^\pw$ with $\pw\approx 1$. We then compare this to theoretical models. We examine a large class of light scalar dark matter models, governed by some potential $V$. For simplicity, we take the scalar to be complex with a global $U(1)$ symmetry in order to readily organize solutions by a conserved particle number. However, we expect our central conclusions to persist even for a real scalar, and furthermore, a complex scalar matches the behavior of a real scalar in the non-relativistic limit which is the standard regime of interest. For any potential $V$, we find the relationship between $\rho_c$ and $R_c$ for ground state solutions is always in one of the following regimes: (i) $\pw\gg1$, or (ii) $\pw\ll 1$, or (iii) unstable, and so it never matches the data. We also find similar conclusions for virialized dark matter, more general scalar field theories, degenerate fermion dark matter, superfluid dark matter, and general polytropes. We conclude that the solution to the core-cusp problem is more likely due to either complicated baryonic effects or some other type of dark matter interactions.
\end{abstract}

\maketitle

\section{Introduction}\label{Introduction}

The distribution of matter in galaxies has been a controversial subject for some time. The vanilla dark matter models (modeled by classical particles with only gravitational interactions) lead to an NFW type profile \cite{Navarro:1995iw}, which appears to match the density profile of the halo of a galaxy quite well. On the other hand, the agreement between vanilla dark matter and observations is less clear towards the center of a galaxy. In particular, most galaxies appear to exhibit a ``core" near their center wherein the density profile $\rho(r)$ flattens as $r\to 0$, while the classic NFW profile has a ``cusp" wherein the density profile $\rho(r)$ rises sharply as $r\to 0$. This discrepancy is known as the ``core-cusp" problem. 

The size of a galactic core is typically on the order of a few kpc. So naively this problem could have an obvious solution: as we go towards the center of the galaxy, on the order of kpc, the baryonic density is rather large, so one can imagine that baryonic effects, such as supernovae and other astrophysical processes, smooth out the center of the galaxy producing a core. However, this candidate explanation is non-trivial to implement for the following reason: the presence of the core appears to persist for small (dwarf) galaxies that are dark matter dominated and baryon deficient. So it is not clear how baryonic feedback effects can solve this puzzle; we shall return to this possibility in the discussion section.

A popular alternate explanation has been put forward for quite some time in which non-vanilla dark matter properties are invoked to smooth out the cores of galaxies. One such popular example is self-interacting dark matter, wherein massive dark matter particles undergo self-scattering with a mean free path $\lambda_{MFP}=1/(n\,\sigma)=m/(\rho\,\sigma)$ with a rather large scattering cross-section $\sigma$. Under come conditions, this can plausibly lead to a core \cite{Spergel:1999mh}. However, such large scattering cross-sections may be at odds with other dark matter measurements, such as observations of the bullet cluster, which puts an upper bound on the dark matter scattering cross-section to mass ratio of $\sigma/m<\mathcal{O}(1)\,\mbox{cm}^2/\mbox{g}$ \cite{Robertson:2016xjh}. If we take a representative core density of $\rho=0.1\,M_\odot\,\mbox{pc}^{-3}$ (see Fig.~\ref{FigureData}), we obtain a mean free path lower bound of $\lambda_{MFP}>\mathcal{O}(50)$\,kpc, which is somewhat larger than the corresponding observed core size of $\approx1$\,kpc; we shall also return to this in the discussion section.

Another popular proposal has been put forward in which the galactic core may be due to the quantum nature of the dark matter particles. In particular, if the dark matter particles are bosons, they can be extremely light, with a huge occupancy number. One of the motivations for this comes from string theory, in which it is plausible that a typical compactification includes axions with exponentially small masses \cite{Arvanitaki:2009fg}. In this case, the de Broglie wavelength $\lambda = h/(m\,v)$ can be of astrophysical size, which is sometimes called ``fuzzy dark matter" \cite{Hu:2000ke}. This gives rise to a kind of ``quantum pressure" that prevents the center of the galaxy from becoming arbitrarily dense, leading to a type of core. A typical mass of $m\sim 10^{-22}$\,eV is often invoked to produce cores of size $\sim$\,kpc. Note that in this regime, the particle's occupancy number must be huge in order for this type of dark matter to be most of the galactic mass. So the theory is well described by classical field theory. (See Ref.~\cite{Hertzberg:2016tal} for a rigorous explanation of why classical field theory provides an accurate description of the dynamics in this regime, despite claims to the contrary in Ref.~\cite{Sikivie:2016enz}.) Other interesting consequences of this proposal are studied in Ref.~\cite{Hui:2016ltb}.

In this work we critically examine this proposal. Our primary motivation is the following: if indeed the large de Broglie wavelength of these ultra-light bosons is responsible for the presence of the core of a galaxy, then it should self consistently explain the core of many other (if not all) galaxies. While one can always fix parameters, say the mass of the dark matter, so that the relationship between core density $\rhoc$ and core radius $\rc$ works for one galaxy, it needs to then correctly predict this relationship for other galaxies; some data is given in Fig.~\ref{FigureData}. Here we show that if the bosons organize into a type of Bose-Einstein condensate, and consequently occupy their ground state configuration near the center of a galaxy, then we can compute this relationship for all galaxies. We find that the resulting relationship between $\rhoc$ and $\rc$ does not match the data for a large family of scalar dark matter models, including ``fuzzy dark matter" in which self-interactions are assumed negligible and for more general scalar field models in which self-interactions are important. We extend these results to scalar dark matter that has not gone into the ground state, but has virialized in a more conventional sense. Finally, we extend our results to even more general scalar field models, degenerate fermions, superfluids, and general polytropes.

Our paper is organized as follows:
In Section \ref{GalacticData} we present some galactic data which indicates $\rhoc\propto 1/\rc$.
In Section \ref{ScalarModels} we present the family of scalar dark matter models that we analyze.
In Section \ref{BoseCondensate} we outline the form of ground state solutions that we are interested in.
In Section \ref{NumericalSolution} we present the numerical solution to these models for a range of parameters.
In Section \ref{AnalyticalApproximation} we describe a simple variational technique to capture the qualitative behavior of the solutions.
In Section \ref{Virialized} we discuss the possibility of non ground state behavior.
In Section \ref{OtherModels} we discuss various other scalar field and fermionic models.
Finally, in Section \ref{SummaryOutlook} we summarize our results and mention future directions.

\section{Galactic Data}\label{GalacticData}

\begin{figure}[t] 
\includegraphics[width=\columnwidth]{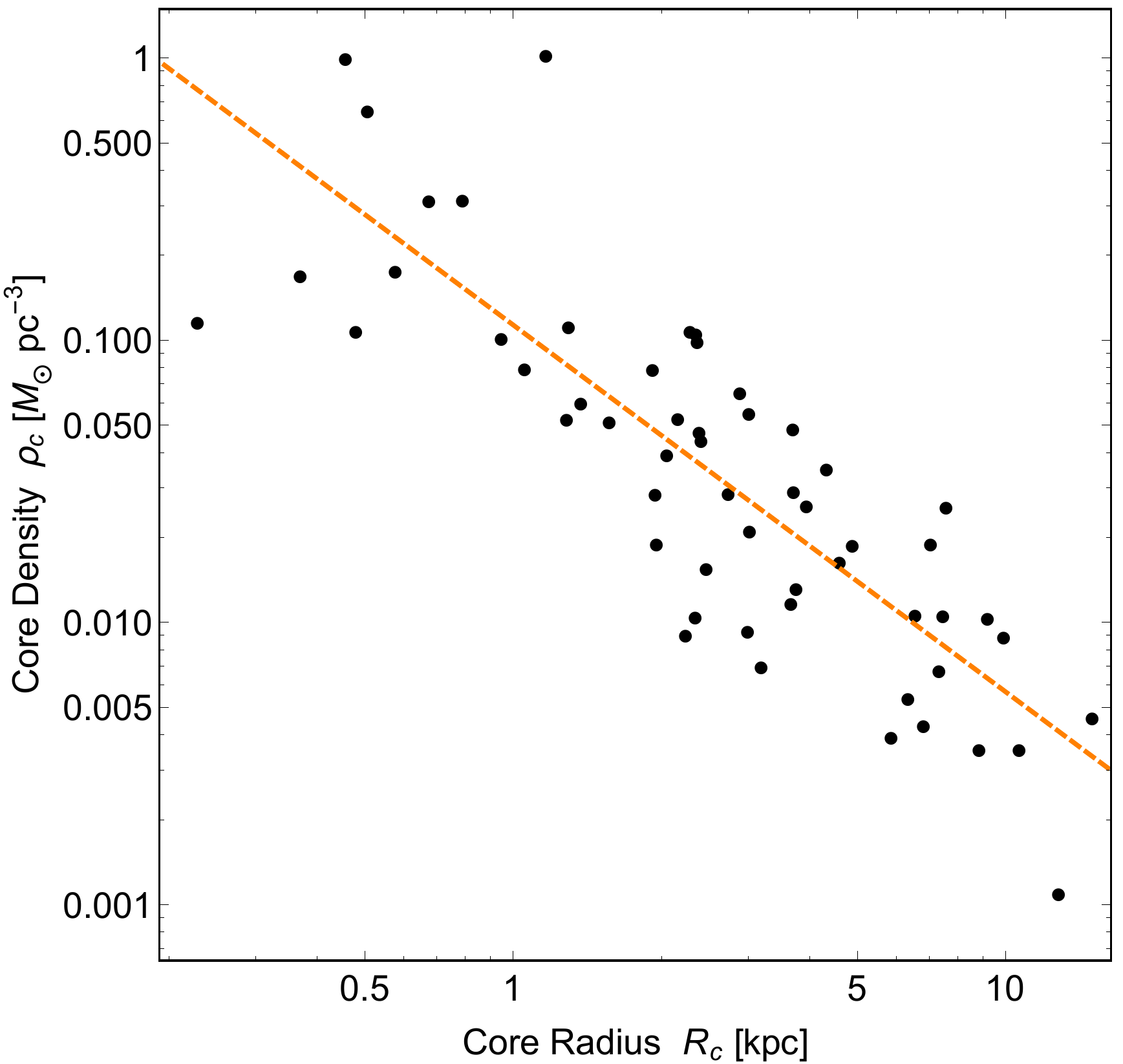} 
\caption{Core density $\rho_c$ versus core radius $R_c$ for a range of galaxies. Black dots are data taken from Ref.~\cite{Rodrigues:2017vto}. Orange dashed curve is the best fit power law curve $\rhoc\propto 1/\rc^\pw$ with $\pw =1.3$ for this data set, while we find $\beta\approx 1$ more generally.}
\label{FigureData}
\end{figure}

The halos of galaxies are dominated by dark matter, which organize into the famous NFW profile \cite{Navarro:1995iw}, in which the density falls off as $\sim1/r^3$ at large radii and rises as $\sim1/r$ at small radii. While this matches data quite well for radii much bigger than $\sim$\,kpc, it appears to fail on scales $r\lesssim$\,kpc. In particular, typical galaxies appear to exhibit a core where the density approaches a constant as $r\to0$. In the vicinity of the core, a convenient density profile is the following functional form \cite{Oh:2015xoa}
\beq
\rho(r) = {\rho_c\over 1+r^2/\rc^2},
\label{rhoprofile}\eeq
where the core density $\rhoc$ is taken to be the central density and the core radius $\rc$ is taken to be the radius at which the density has dropped to half its central value. In Fig.~\ref{FigureData} we plot core density $\rhoc$ versus core radius $\rc$ for a range of different galaxies from Ref.~\cite{Rodrigues:2017vto}. (In fact this reference used the so-called Burkert profile $\rho(r)=\rhoc/(1+r/R_B)(1+r^2/R_B^2)$ in which the core radius $R_B$ is when the density has dropped to a quarter of its central value; so we have re-scaled by a factor of $\rc\approx 0.54R_B$ accordingly.) Their data $\{\rc,\rhoc\}$ comes from measuring rotation curves. Although there is significant scatter in the data, an overall trend can be clearly seen. By parameterizing the relationship between core density and core radius as a power law
\beq
\rhoc\propto {1\over\rc^\beta},
\eeq
we find that the best fit value for the exponent for this particular data set is $\pw=1.3$. We have also examined other data sets, including faint galaxies \cite{Oh:2015xoa}, finding $\beta=0.9$, etc. In general we find that $\beta\approx 1$. We note that this roughly holds for both small (dwarf) and large galaxies, including galaxies that are dominated by dark matter.

\section{Scalar Models}\label{ScalarModels}

Our task then is to see whether the above behavior can be reproduced by extremely light bosons. As a reasonable model for light bosonic dark matter, we take the dark matter particle to be spin 0. One could imagine studying a massive spin 1 particle, though we expect qualitatively similar behavior. For simplicity, we will focus in this paper on a complex scalar $\f$ whose dynamics is organized by a global $U(1)$ symmetry $\f\to\f\, e^{i\theta}$. (Other work on complex scalar dark matter includes Ref.~\cite{Li:2013nal}.) We do not have a physical motivation for this ad hoc symmetry, we only introduce this to simplify the analysis. In particular, the global symmetry leads to a conserved particle number $\N$, which we can use to organize solutions. More precisely, we will focus on ground state solutions at fixed number $\N$. If instead we studied a real scalar field, there would technically be no conserved particle number to organize solutions, and so the only true ground state solutions would be vacuum solutions. So for real scalars with self-interactions, one is forced to consider particle number changing processes, whose time dependence can be complicated in general. However, this problem only arises at large field values. At small field values, particle number changing processes are suppressed, and the real scalar field evolution matches the complex scalar field evolution. We shall return to the difference between complex and real scalar field dynamics in the next section and the discussion.

We assume this scalar $\f$ is minimally coupled to Einstein gravity with a canonical kinetic term and allow for self-interactions from a potential $V$. The action is then (signature + - - -, units $\hbar=c=1$)
\bea
S=\int\! d^4x\sqrt{|g|}\left[{\mathcal{R}\over16\pi G} + |\partial\f|^2 - V(|\f|)\right].
\eea
By varying the action, we obtain the standard Einstein equations for gravity, as well as the following equation of motion for $\f$
\beq
\Box\f\equiv |g|^{-1/2}\,\partial_\mu\!\left(\sqrt{|g|}\,g^{\mu\nu}\partial_\nu\f\right)=-\partial_{\f^*} \!V.
\label{BoxPhi}\eeq

The primary choice to make is the potential $V$. Since we know cold dark matter (described by non-relativistic particles) works very well on large scales, we demand that the potential is quadratic $V = m^2|\f|^2+\ldots$ around its minimum. At large field values, one is allowed to consider various possibilities for the potential. For simplicity, we consider potential functions that are monotonically increasing as we increase $\Phi$. Otherwise we would have the complication of potentials with multiple minima. This could lead to topological defects that are currently unobserved and so we shall avoid this possibility here. 

A family of potential functions that captures a range of qualitatively different behavior is the following
\bea
V(|\f|) = {m^2\F^2\over\ex}\left(\left(1+{|\f|^2\over\F^2}\right)^{\!\!\ex}-1\right),
\eea
where $\F$ is some mass scale that represents the cutoff of the effective theory and $\ex>0$ is a positive exponent. Note that if we Taylor expand the potential around small $\f$, we have
\beq
V(|\f|)=m^2|\f|^2+\g|\f|^4+\ldots,
\label{Vexpand}\eeq
where 
\beq
\g\equiv{m^2(\ex-1)\over 2\F^2}.
\eeq
So if $\ex=1$ we have no self-interactions, if $\ex>1$ we have repulsive self-interactions ($\g>0$), and if $\ex<1$ we have attractive self-interactions ($\g<0$). The case of $\alpha<1$ is therefore representative of (i) typical axion models (albeit axions are real scalars $\phi$) in which the canonical single instanton potential $V(\phi)=m^2\Fa^2(1-\cos(\phi/\Fa))$ gives the small field expansion $V(\phi)=m^2\phi^2/2-m^2\phi^4/(24\Fa^2)+\ldots$ and (ii) axion-monodromy models in which the above potential is indicative \cite{McAllister:2008hb}. For a possible construction of a repulsive ($\alpha>1$) light dark matter model see \cite{Fan:2016rda}.

\section{Bose-Einstein Condensate}\label{BoseCondensate}

The above action's global $U(1)$ symmetry gives rise to the following conserved current associated with particle number
\beq
J^\mu = i\sqrt{|g|}\,g^{\mu\nu} (\f^*\,\partial_\nu\f-\f\,\partial_\nu\f^*),
\eeq
with $\partial_\mu J^\mu=0$ and a corresponding conserved particle number
\beq
\N=\int d^3x\,n({\bf x},t),\,\,\,\, \mbox{with}\,\,\,\, n({\bf x},t)\equiv J^0({\bf x},t)
\label{Number}\eeq
the local number density. 

Since particle number is conserved and since we have bosons at high occupancy, the system can in principle organize into a Bose-Einstein condensate. This occurs if the system has sufficiently fast interactions to re-organize towards the ground state. As discussed in Refs.~\cite{Sikivie:2009qn,Erken:2011dz,Guth:2014hsa} the gravitational interaction rate for a typical mode $k$ for an initially messy field configuration can be estimated as
\beq
\Gamma\sim{8\pi G\,m^2\,n\over k^2}.
\eeq
At first sight this appears to be tiny for very light fields due to the $m^2$ factor. However, this is incorrect for the following reasons: (i) in order for $\f$ to be the bulk of the dark matter in the galaxy, then we need $\rho=m\,n$ fixed to the observed galactic density, and (ii) we can re-write the wavenumber as $k=m\,v$, where $v$ is a virial speed in the galaxy. This gives $\Gamma\sim8\pi G\rho/(m\,v^2)\propto 1/m$ and so it is evidently becoming large for very light fields. A representative value from Fig.~\ref{FigureData} is $\rho=0.1\,M_\odot\,\mbox{pc}^{-3}$ with corresponding virial speed $v\approx 10^{-4}\,c$ (using Eq.~(\ref{virialspeed}) with $R=$\,kpc). Then if we take $m= 10^{-22}$\,eV, we obtain $1/\Gamma\approx 5$\,Myr, which is smaller than a typical galactic age (in fact this time-scale can be further reduced due to self-interactions). This suggests there may be enough time for gravitational interactions to organize the field towards a Bose-Einstein condensate, which will be our assumption going forward for most of this paper. However, see Section \ref{Virialized} for a discussion of the case in which this assumption is not satisfied.

The Bose-Einstein condensate is a configuration that minimizes the energy at fixed particle number. This comes from extremizing the free energy
\beq
E-\mu\,N,
\eeq
where $E$ is the energy and $\mu$ is a chemical potential. It can be readily shown that this is extremized for field configurations with the following simple time and space dependence
\beq
\f({\bf x},t) = {\fr(r)\over\sqrt{2}}\,e^{-i\,\omega\,t},
\label{SimpleTS}\eeq
where $\omega=\mu$ is the frequency of oscillation of the field in the complex plane. Here $\fr(r)$ is some real function of radius only, as the ground state selects out a spherically symmetric configuration. Since we will extremize the free energy, we can in general find both local minima (stable solutions) and local maxima (unstable solutions).

Note that if we were to consider our starting field $\f$ to be a real scalar, then in the non-relativistic limit, we would simply add to the right hand side of Eq.~(\ref{SimpleTS}) the complex conjugate. The equation of motion for $\fr(r)$ would essentially remain the same, up to some trivial re-scalings by $\mathcal{O}(1)$ factors. (See Ref.~\cite{Namjoo:2017nia} for a systematic treatment of corrections to the non-relativistic theory of a real scalar.)

\section{Numerical Solution}\label{NumericalSolution}

Here we would like to describe our numerical recipe and numerical results for the above set of models.

\subsection{Setup}

For spherically symmetric static (ground state) solutions, we can write the space-time metric without loss of generality as \cite{Friedberg:1986tp}
\beq
ds^2= e^{2\mphi(r)}\mS(r) dt^2-\mS(r)^{-1}dr^2-r^2 d\Omega^2,
\eeq
where $\mphi$ and $\mS$ are constrained variables which are functions of radius $r$. Here $\mS(r)=1-2GM(r)/r$, with $M(r)$ the enclosed energy up to radius $r$. The total energy is
\beq
E = M(r\to\infty).
\eeq
The total energy density is given by a sum over kinetic $T$, gradient $W$, and potential $V$ energy densities, as follows
\beq
\rho(r) = T(r)+W(r) + V(r),
\eeq
where
\beq
T={1\over2}e^{-2\mphi}\mS^{-1}\omega^2\,\fr^2,\,\,\,\, W={1\over2}\mS\,\fr'^2.
\eeq
Here a prime $'$ denotes a derivative with respect to radius. The enclosed energy $M$ and the metric function $\mphi$ are then obtained from the following pair of 1st order ODEs
\bea
M' & = & 4\pi r^2\rho,\\
\mphi' & = & 8\pi G r\mS^{-1}(T+W).
\eea

The equation of motion for the scalar field Eq.~(\ref{BoxPhi}) in this spherically symmetric static configuration becomes the following 2nd order ODE
\beq
e^{-2\mphi}\mS^{-1}\omega^2\fr+\mS \fr''+\left({1+\mS\over r}-8\pi G r V\right)\!\fr'=\partial_\fr V,
\eeq
and the conserved particle number Eq.~(\ref{Number}) is given by the following integral
\beq
N = 4\pi\,\omega\! \int_0^\infty \!dr \, r^2\,e^{-\mphi}\mS^{-1}\fr^2.
\eeq

\begin{figure}[t] 
\includegraphics[width=\columnwidth]{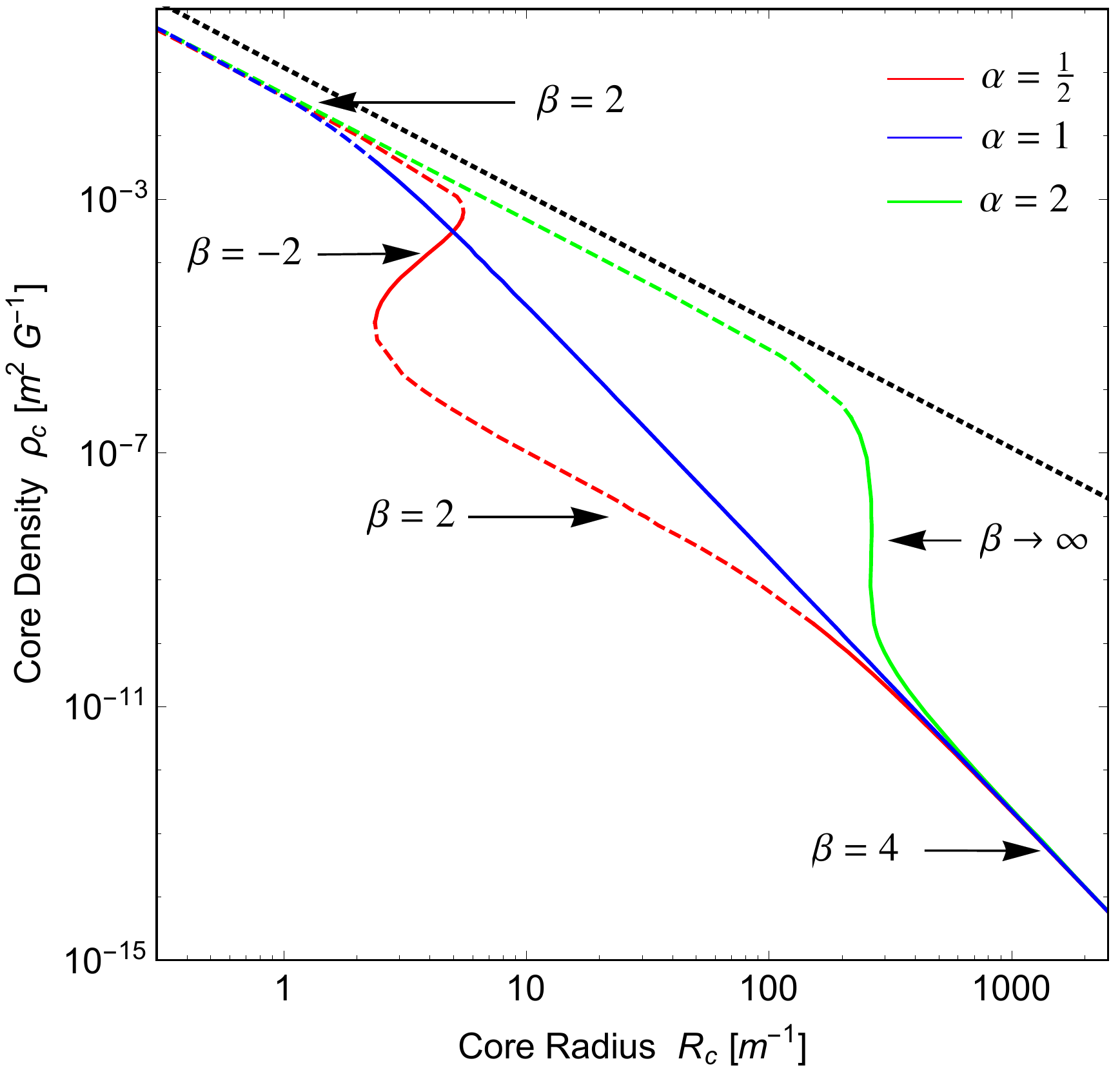} 
\caption{Core density $\rhoc$ versus core radius $\rc$ for different exponents $\ex$ in the potential functions $V=m^2F^2((1+|\f|^2/\F^2)^\ex-1)/\ex$. Here we have scaled out the constants $m$ and $G$, and chosen $F=10^{-3}/\sqrt{G}$ ($\approx 10^{16}$\,GeV). The lower red curve is for $\ex=1/2$ (attractive self-interaction), the middle blue curve is for $\ex=1$ (no self-interaction), and the upper green curve is for $\ex=2$ (repulsive self-interaction). The solid branches represent stable solutions (true ground states), while the dashed branches represent unstable solutions. The different branches all asymptote to power law behavior $\rhoc\propto 1/\rc^\pw$ with the value of $\pw$ labelled for each branch. The dotted black line indicates the boundary of the black hole regime, above which the scalar field would be trapped inside its own Schwarzschild radius; since the cores of galaxies are in the weak gravity regime, the important region of this plot is the part well below the dotted black line.}
\label{FigureTheory}
\end{figure}

\subsection{Results}

We have solved these equations numerically for a range of values of the exponent $\ex$ in the potential $V$. For each field amplitude at the center $\phi(0)$, we numerically determine the corresponding frequency $\omega$ that gives rise to a localized solution with no nodes. This involved repeated trials of different $\omega$ until the correct one is obtained with high accuracy. We then determined the core density and radius for this solution. This process is then repeated for many different field amplitudes. 

Our results are summarized in Fig.~\ref{FigureTheory}, where we plot the core density (defined by $\rhoc=\rho(0)$) versus the core radius (defined implicitly by $\rho(\rc)=\rho(0)/2$). We have scaled out the mass $m$ and gravitational constant $G$ by measuring $\rc$ in units of $m^{-1}$ and $\rhoc$ in units of $m^2\,G^{-1}$. Then the physics is controlled by the one residual scale $F$, which we have chosen to be $F=10^{-3}/\sqrt{G}$ ($\approx 10^{16}$\,GeV) in this plot. We have shown 3 representative values of the exponent $\ex$: $\ex=1/2$ (attractive self-interaction), $\ex=1$ (no self-interaction), and $\ex=2$ (repulsive self-interaction). 

The results in Fig.~\ref{FigureTheory} show that the solution is somewhat complicated, undergoing multiple branches for each choice of $\ex$. As we will explain in the next section, each branch can be described by a power law $\rhoc\propto 1/\rc^\pw$ for some value of the power $\pw$. Furthermore, some branches are stable (solid lines) and some branches are unstable (dashed lines). As indicated in the plot, there is no regime in which $\pw\approx 1$. There are branches where $\pw=2$, which is not so far off the observational data, however those are always unstable solutions. 

To summarize, we find that in all branches we have (i) $\pw\gg 1$, or (ii) $\pw\ll1$, or (iii) unstable solutions. So none of these branches is compatible with the galactic data (see Fig.~\ref{FigureData}). We find that this result persists for any choice of the exponent $\ex$ and any choice of the parameter $F$, and we believe this persists for general potential functions $V$ (some generalizations are in Section \ref{PotentialCorrections}).

\section{Analytical Approximations}\label{AnalyticalApproximation}

In this section we would like to provide an analytical understanding of the above numerical results. 

\subsection{Non-Relativistic Regime}

Firstly, the standard regime of interest in light scalar dark matter models is the non-relativistic regime, where fields are small allowing the potential to be approximated as $V\approx m^2|\f|^2+g|\f|^4$, gravity is given by the weak field Newtonian limit, and gradient energies are small compared to kinetic energies. In this non-relativistic limit the above total energy reduces to the following sum of mass-energy, gradient energy, self-interaction energy, and gravitational energy, respectively \cite{Schiappacasse:2017ham,Visinelli:2017ooc}
\bea
E &=& \N\,m +4\pi\!\int_0^\infty \!dr\,r^2\left[{\fr'^2\over2} +{\g\over 4}\fr^4\right]\nonumber\\
 & - & {Gm^4\over2}(4\pi)^2\!\int_0^\infty \!dr\,r^2\! \int_0^\infty \!dr'\,r'^2\,{\fr(r)^2\fr(r')^2\over r_>},\,\,\,\,\,\,
\label{energynonrel}\eea
where $r_>$ is the greater of the pair $\{r,r'\}$, and the particle number in this limit is simply 
\beq
\N=4\pi\,m\int_0^\infty \!dr\,r^2\,\fr^2.
\eeq
Note that these equations are identical to the non-relativistic limit of a real scalar; which is in fact the classic form of fuzzy dark matter, including ultra-light axions.
Following Ref.~\cite{Schiappacasse:2017ham}, we take a simple exponential ansatz for the spatial profile as $\phi(r)=\sqrt{\N/(\pi\,m\,\rc^3)}\,\exp(-r/\rc)$ (here $\rc$ is the radius at which the density is $e^{-2}$ of its central value, which can be trivially re-scaled to give the radius at which the density is $1/2$ of its central value if desired). Then we obtain the following energy
\beq
E = \N\,m+{\N\over 2\,m\,\rc^2} + {\g\,\N^2\over32\,\pi\, m^2\,\rc^3} -{5\,G\,m^2\,\N^2\over 16\,\rc}.
\label{EnonRel}\eeq
Static solutions arise from extremizing this energy with respect to core radius $\rc$ at fixed particle number $\N$.

For large radii, we need to balance the 2nd (gradient) and 4th (gravitational) energies in Eq.~(\ref{EnonRel}). This requires $\rc\propto 1/\N$. So then $E\approx\N\,m\propto1/\rc$. Then using $\rhoc\sim E/\rc^3$, this implies
\beq
\rhoc\propto {1\over\rc^4},\,\,\,\,\mbox{large}\,\,\rc\,\,\,\,(\mbox{stable}),
\eeq
where we have indicated that this solution is stable, since it is a local minimum of the above energy function. This accounts for the lower right $\pw=4$ region of Fig.~\ref{FigureTheory}, which is obeyed by all potential models, independent of $\ex$, in this gravitationally dominated limit. This is an important result: when ultra-light axions are in the standard regime in which their behavior is governed by Newtonian gravity, the model predicts $\rho\propto 1/\rc^\pw$ with $\pw=4$, rather than $\pw\approx 1$, which is favored by the data.

For repulsive self-interactions ($\ex>1$ and $\g>0$) another branch of solutions emerges as we go to higher particle number $\N$. In this case the 3rd (self-interaction) and 4th (gravitational) energies in Eq.~(\ref{EnonRel}) can balance each other. Since they both scale as $\N^2$, this requires $\rc=\mbox{const}$, leading to a vertical line, which we can express heuristically as
\beq
\rhoc\propto {1\over\rc^\infty},\,\,\,\,\mbox{repulsive only}\,\,\,\,(\mbox{stable}).
\eeq
This accounts for the vertical branch of the repulsive (green) case in Fig.~\ref{FigureTheory}.

For attractive self-interactions ($\ex<1$ and $\g<0$) a different type of solution emerges as we decrease radius $\rc$. In this case the 2nd (gradient) and 3rd (self-interaction) energies in Eq.~(\ref{EnonRel}) can balance each other. This requires $\rc\propto\N$. So then $E\approx\N\,m\propto\rc$, which implies
\beq
\rhoc\propto {1\over\rc^2},\,\,\,\,\mbox{attractive only}\,\,\,\,(\mbox{unstable}).
\eeq
Note that since this only leads to an extremum if $g<0$, this gives rise to a local maximum of the above energy function, leading to an unstable branch. This accounts for the centrally located dashed (red) branch in Fig.~\ref{FigureTheory} with $\pw=2$.

\subsection{Large Scalar Field Regime}

There are two basic ways in which the above non-relativistic theory can breakdown. The first is if the field amplitude becomes sufficiently large that the potential $V$ is no longer dominated by its mass term, meaning frequencies are no longer near $m$, as we now explore.

At large field amplitudes, $|\f|=\fr/\sqrt{2}\gg\F$, the potential can be approximated as
\beq
V\approx \coef\, \fr^{2\ex},
\eeq
where $\coef\equiv m^2\F^{2-2\ex}/(2^{\ex}\ex)$. As above this can permit a regime in which the self-interaction is dominant and gravity is negligible, however we need to now operate fully relativistically. In this case, the total energy becomes
\beq
E = 4\pi\!\int_0^\infty\!dr\,r^2\left[ {\omega^2\fr^2\over2}+{\fr'^2\over2}+\coef\,\fr^{2\ex}\right],
\eeq
with particle number $\N=4\pi\,\omega\int_0^\infty \!dr\,r^2\,\fr^2$. By again taking a simple exponential ansatz for the spatial profile as $\phi(r)=\sqrt{\N/(\pi\,\omega\,\rc^3)}\,\exp(-r/\rc)$, we obtain the energy
\beq
E = {\omega\,\N\over 2}+{\N\over 2\,\omega\,\rc^2}+\tilde\xi\,{\rc^{3-3\ex}\N^\ex\over\omega^\ex},
\eeq
where $\tilde\xi\equiv\pi^{1-\ex}\xi/\ex^3$. By extremizing $E$ with respect to both $\omega$ and $\rc$, one can show that solutions only exist for $\ex<1$, $\omega\sim1/\rc$, and $\rc\propto\N^{(1-\ex)/(4-2\ex)}$. So then $E\sim \N/\rc\propto \rc^{(3-\ex)/(1-\ex)}$, which implies
\beq
\rhoc\propto \rc^{2\ex\over1-\ex},\,\,\,\,\mbox{attractive only}\,\,\,\,(\mbox{stable}),
\eeq
which can be readily shown to be a stable branch. This accounts for the upper (red) attractive case in Fig.~\ref{FigureTheory}, which for $\ex=1/2$, gives $\pw=-2$. These types of solutions are often referred to as ``Q-balls" in the literature \cite{Coleman:1985ki}. Since $0<\ex<1$ for these solutions to exist, they always give a core density that grows with radius, in clear contradiction to the galactic data.

\subsection{Strong Gravity Regime}

The second way in which the non-relativistic theory can breakdown is when we enter the strong gravity regime with gravitational potential $G\,M/r=\mathcal{O}(1)$. Such a regime is invariably associated with orbital speeds that are a significant fraction of the speed of light, which cannot possibly account for the behavior near the cores of galaxies on scales $\sim$\,kpc, which are measured to have $v\sim 10^{-4}\,c-10^{-3}$\,c. Nevertheless we include this here for completeness. 

As we explore solutions with smaller and smaller radii and higher densities, the core radius $\rc$ gets closer and closer to its Schwarzschild radius $R_S=2GM$; the radius of a spherically symmetric black hole. To indicate this region, let us define a critical density
\beq
\rho_S\equiv {3\,M\over 4\,\pi\, R_S^3}={3\over8\,\pi\,G\,R_S^2}.
\eeq
At a fixed radius, any density above this critical value will be a region that is trapped inside its own Schwarzschild radius and will collapse to a black hole. We have indicated this critical density by a dotted black line in Fig.~\ref{FigureTheory}. 

Note that all scalar field static solutions stay below, but become close and parallel to, this critical density in the upper left region of the figure. These solutions are evidently not black holes, but are in the strong gravity regime with a radius that is only a factor of a few larger than their Schwarzschild radius. Hence these solutions are characterized by the same power law as the black hole, namely
\beq
\rhoc\propto {1\over\rc^2},\,\,\,\,\mbox{small}\,\,\rc\,\,\,\,(\mbox{unstable}).
\eeq
Such solutions are unstable as they can collapse to a black hole under the appropriate perturbation.

\section{Virialized Behavior}\label{Virialized}

In the analysis above we have assumed that the scalar field has condensed into its ground state at the core of a galaxy. Since the gravitational thermalization rate $\Gamma\sim 8\pi G\rho/(m\,v^2)$ gave a value $1/\Gamma\sim 5$\,Myr for typical input parameters, this assumption seems at least plausible. However, we do not have a proof that this would happen, so it is useful to consider the case in which the field has yet to fall into its ground state.

\begin{figure}[t] 
\includegraphics[width=\columnwidth]{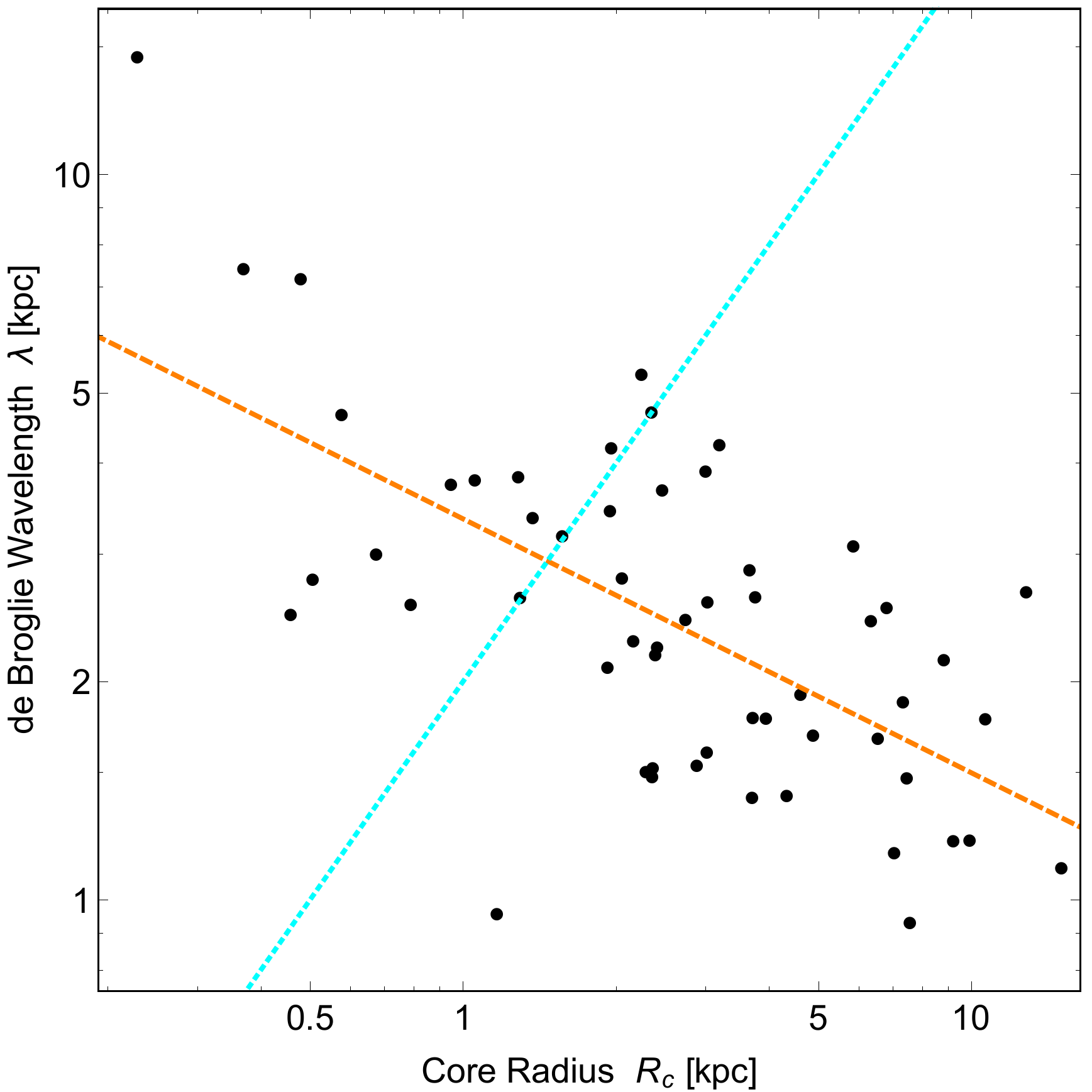} 
\caption{de Broglie wavelength $\lambda=h/(m\,v)$ versus core radius $R_c$ for a range of galaxies in the fuzzy dark matter hypothesis. Black dots correspond to the data from Ref.~\cite{Rodrigues:2017vto} that we used in Fig.~\ref{FigureData}. We have taken the velocity $v$ that determines the de Broglie wavelength to be the virial speed corresponding to the core density $\rhoc$ for that radius $\rc$, and we have taken the dark matter particle mass to be $m=10^{-22}$\,eV for the sake of illustration (since $\lambda\propto 1/m$, other values of $m$ involves a simple re-scaling of the vertical axis). Dashed orange curve is the corresponding best fit power law $\lambda\propto 1/\rc^{1-\pw/2}$ from Fig.~\ref{FigureData} with $\pw =1.3$ for this data set. The dotted cyan curve is $\lambda=2\,\rc$; for points that lie well above this line, the theoretical model is unphysical, and for points that lie well below this line, the core seems to require some alternate explanation.} 
\label{FigureDataDB}
\end{figure}

In this case we still expect the field's velocity distribution to virialize. Let us focus here on the simplest version of ultra-light scalar dark matter in which self-interactions are negligible and the dynamics is governed by Newtonian gravity. The virialized speed at the core radius is
\beq
v=\sqrt{G\,M_{c}\over \rc},
\label{virialspeed}\eeq
where $M_{c}$ is the enclosed mass up to radius $\rc$. In the vicinity of the core, we take the density profile to be given by the fiducial form in Eq.~(\ref{rhoprofile}). Integrating this gives the core mass as 
\beq
M_c = (4-\pi)\pi\,\rhoc\,\rc^3.
\label{coremass}\eeq
Now a non-relativistic quantum particle has a de Broglie wavelength set by its characteristic speed $v$ as
\beq
\lambda = {h\over m\,v}.
\eeq
By using Eqs.~(\ref{virialspeed}, \ref{coremass}) this expression gives the de Broglie wavelength in the vicinity of the core of a galaxy as a function of core density and radius. This is given in Fig.~\ref{FigureDataDB} as a function of core radius for a range of observed galaxies, where we have used the corresponding core density data from Fig.~\ref{FigureData}. We chose a particle mass of $m=10^{-22}$\,eV for illustrative purposes.
We have also included the corresponding best fit curve (dashed orange), which is $\lambda\propto 1/\rc^{1-\pw/2}$ with $\pw=1.3$ for this data set.

Note that the data indicates that the de Broglie wavelength is a decreasing function of the galactic core radius. However, this appears to go in the opposite direction to the idea behind the ultra-light or ``fuzzy" dark matter proposal. To illustrate this we have also plotted as the dotted cyan curve $\lambda=2\,\rc$. Any galaxies that lie well above this line would have a core diameter that is much smaller than the corresponding scale over which the particles are localized, which we consider to be an unphysical prediction of the model. On the other hand, any galaxies that lie well below this line have a core diameter that is much bigger than the size of the particle's wave-packet. While this latter scenario can be perfectly physical, it begs the question as to what is then actually responsible for the large core size, since the proposal of fuzzy dark matter is that the core arises from the particle's de Broglie wavelength itself. In fact to match $\lambda=2\,\rc$, the data should follow $\rhoc\propto1/\rc^4$. Since the orange curve (best fit data) and cyan curve (theoretical prediction) are essentially orthogonal to each other, it disfavors this proposal.

\section{Other Models}\label{OtherModels}

In this Section we generalize our results to a range of other models, including more general scalar theories, fermions, superfluids, and general polytropes.

\subsection{Other Scalar Field Theories}

\subsubsection{Real Scalars}

An important subject is that of a real scalar field. However, we expect similar behavior to the case of the complex field studied here. In fact in the non-relativistic limit, both theories obey the same equations of motion. Hence the standard regime of the ultra-light axion scenario is fully encompassed by our analysis here. However, for large field amplitudes, there can be differences. In particular, there can be particle number changing processes allowed. It appears unlikely that such processes could at all help to explain galactic cores; if anything, such behavior would limit the stability of such cores, making it even less likely to produce a consistent model.

\subsubsection{Kinetic Corrections}

Other possibilities are to include corrections to the action, including higher order kinetic terms, such as 
\beq
\Delta\mathcal{L}= \gam\,|\partial\f|^4+\ldots,
\eeq
for a complex field or a real field ($\gam$ is some coupling). However, in the non-relativistic limit, this introduces the correction 
\beq
\Delta\mathcal{L} = \gam\,m^4\,|\f|^4+\ldots,
\eeq
which is merely a correction to the quartic term in the potential function $V$, with $\Delta\g=-\gam\,m^4$. Hence in this non-relativistic regime (which is the primary regime of interest for galactic cores) this does not introduce any new behavior that is not already captured by the earlier analysis in this paper. There may be new behavior in the relativistic regime, however, which we will not pursue here.

\subsubsection{Potential Corrections}\label{PotentialCorrections}

Another possibility is to generalize the small field expansion of the potential $V$ in Eq.~(\ref{Vexpand}) to
\beq
V=m^2|\f|^2+\g\,|\f|^{2\expo}+\ldots.
\eeq
Previously we considered $\expo=2$, which is the standard form expected of the leading self-interaction term. One is allowed to consider other values of $\expo$ for the sake of generality. We shall focus here on the non-relativistic limit. In this case one may simply assume that $\Delta V=\g\,|\f|^{2\expo}$ is present in some non-relativistic effective theory, regardless of its origin. From this point of view, we may even allow $\expo$ to be a fractional power. In the non-relativistic limit, the energy becomes a simple generalization of Eq.~(\ref{energynonrel}) by replacing the term $\g\,\phi^4/4\to\g\,\phi^{2\expo}/2^\expo$. Once again we have two scenarios in which the self-interaction term can be important. The first is if it balances the gradient energy, which we shall study here. The second is if it balances the gravitational energy, which we shall study in Section \ref{Polytrope}.

The self-interaction and gradient energies together scale as
\beq
\Delta E=c_1{\g \,N^\expo \over m^\expo\,\rc^{3(\expo-1)}} + c_2{N\over m\,\rc^2},
\eeq
where $c_{1,2}$ are positive $\mathcal{O}(1)$ numbers. Balancing these two terms and using $\rhoc\propto N/\rc^3$ leads to
\beq
\rhoc\propto {1\over\rc^\pw}\,\,\,\,\mbox{with}\,\,\,\beta={2\over\expo-1}.
\eeq
For $\g>0$ the existence of a solution requires $\expo<1$ (which would be a strange non-local term in the effective theory) leading to $\pw<0$.
For $\g<0$ a stable solution requires $0<3(\expo-1)<2$ leading to $\pw>3$. In either case it is disfavored by the data.

\subsection{Degenerate Fermions}

Another possibility is to study moderately light fermions, which will undergo Pauli exclusion in the degenerate regime. This can lead to a core at the center of galaxies if the fermion mass is sub-keV \cite{Randall:2016bqw}. The relationship between core density and core radius in this scenario can be estimated as follows: In the degenerate regime $\rhoc=m\,\nc\sim m/\lambda^3\sim m(m\,v)^3$, and virialization implies $v\sim\sqrt{G\mc/\rc}\sim\sqrt{G\,\rhoc\,\rc^2}$. Eliminating $v$ from this pair of equations, we have 
\beq
\rhoc\propto {1\over\rc^6}, 
\label{Fermion}\eeq
which is ruled out by galactic data.

\subsection{Superfluid Dark Matter}

An interesting proposal for the dark matter is that it permits a type of phase transition to a superfluid state that implements a type of MONDian dynamics on galactic scales \cite{Berezhiani:2015bqa}. The effective Lagrangian for a scalar field $\theta$ in this regime is $\mathcal{L}\propto X\sqrt{|X|}$, where $X=\dot\theta-m\,\phi_N-(\nabla\theta)^2/(2\,m)$. This is associated with an equation of state $P\propto\rho^3$ \cite{Berezhiani:2015bqa}, which leads to a core (see next subsection), with relation
\beq
\rhoc\propto \rc^2,
\label{Superfluid}\eeq
which is also disfavored by the data. However, this model involves non-trivial coupling to baryons - such that dark matter somewhat mimics MONDian gravity - which may alter the predictions. See e.g. Ref.\cite{Milgrom:2004ba} for discussions on MOND and its predictions. 

\subsection{General Polytropes}\label{Polytrope}

The above models are examples of systems where the pressure (e.g., from self-interactions or Fermi degeneracy) is simply a function of the density, which is known as a polytrope equation of state
\beq
P=K\,\rho^{1+{1\over \ind}}.
\eeq
Here $\ind$ is known as the polytropic index and $K$ is a constant of proportionality. For $\g|\f|^{2\expo}$ potential corrections we have $\ind=1/(\expo-1)$, for degenerate fermions we have $p=3/2$, and for the above superfluid dark matter model we have $\ind=1/2$. 

In astrophysical systems, this pressure force is used to balance against the gravitational force to achieve some equilibrium configuration (note for ordinary scalar fields, the ``pressure" here refers to self-interactions, and not to be confused with the ``quantum pressure" or gradient energy, which is not of the polytropic form). Since pressure is a force per unit area, this gives rise to a pressure force of characteristic size $F_P\sim P\,r^2\propto \rho^{1+{1\over\ind}}r^2$. On the other hand, gravity is a force of characteristic size $F_G\sim G M^2/r^2\propto \rho^2\, r^4$. In hydrostatic equilibrium these forces balance each other, leading to the following relationship between core density and core radius
\beq
\rhoc\propto {1\over\rc^\pw}\,\,\,\,\mbox{with}\,\,\,\beta={2\,\ind\over\ind-1},
\eeq
which reproduces the fermion result in Eq.~(\ref{Fermion}) for $\ind=3/2$ and the superfluid result in Eq.~(\ref{Superfluid}) for $\ind=1/2$. Stable solutions from this balance between pressure and gravity requires $0<\ind<3$ (and $K>0$). Hence the exponent $\pw$ is constrained to be either $\pw<0$ or $\pw>3$. So once again it can never be close to 1 in order to match the galactic data.

\section{Summary and Outlook}\label{SummaryOutlook}

We have shown that it is very difficult for light dark matter to reproduce the observed relationship between core density and core radius in galaxies, which obeys the rough scaling law $\rhoc\propto 1/\rc^\pw$ with exponent $\pw\approx 1$. 

In particular, we have shown the following: ultra-light scalars, with negligible self-interactions, lead to $\pw=4$ when Newtonian gravity is balanced against ``quantum pressure"; large self-interactions give rise to the wrong $\pw$ for any potential and/or instabilities; the strong gravity regime can lead to instabilities and is in any case strongly disfavored by the data; virialized ultra-light scalars predict the wrong relation between the de Broglie wavelength and core radius; kinetic corrections to the scalar field Lagrangian are redundant with these results in the non-relativistic limit; and any polytrope equation of state leads to instabilities and/or the wrong exponent, including potential corrections, degenerate fermions, and superfluid dark matter. 

Further work would be to generalize this class of theories. In order to have the correct scaling in this context of very light scalars, would appear to involve unusual fractional derivatives, etc, which is associated with non-locality. Such effective theories may be very difficult to reconcile with a range of other observations. Other important work is to perform numerical simulations to determine to what extent the field organizes into its ground state versus other states. 

A natural possibility is to return to heavy dark matter particles that may exhibit more standard interactions. As mentioned in the introduction, if dark matter particles have a large scattering cross section $\sigma$, they will have a finite mean free path $\lambda_{MFP}=1/(n\,\sigma)=m/(\rho\,\sigma)$. Interestingly, if this mean free path sets the size of the core, and if the cross section is velocity independent, then this naturally predicts $\rhoc\propto 1/\rc$. However, there are a range of constraints on these strongly interacting models, such as from bullet cluster observations, halo properties, etc, so this requires careful future analysis. In any case, these types of interactions, or other possible dark matter interactions, are worthwhile exploring as a possible solution to the core-cusp problem.

Finally, one should include baryons into simulations as fully as possible to examine whether this may explain the presence of cores. Since baryons tend to accumulate towards the center of galaxies, this may be very important and indeed may well explain the discrepancies, at least, for largest galaxies. However, galactic cores tend to persist even for smaller galaxies, including those that are rather dark matter dominated, so it is unclear what the final solution will be.

\section*{Acknowledgments}
We would like to thank Sergei Dubovsky, Alan Guth, Lam Hui, and Alex Vilenkin for helpful discussions. MPH is supported by National Science Foundation grant PHY-1720332. MHN is supported in part by the U.S. Department of Energy under grant Contract Number de-sc0012567. HD is supported by the John F. Burlingame Graduate Fellowships in Physics at Tufts University.

\end{document}